# Utilization of Wind Energy at High Altitude*

**Alexander Bolonkin**
V.P. of Consulting and Research Co.
1310 Avenue R, #6-F, Brooklyn, NY 11229, USA
T/F 718-339-4563. aBolonkin@juno.com, aBolonkin@gmail.com, http://Bolonkin.narod.ru

## Abstract

Ground based, wind energy extraction systems have reached their maximum capability. The limitations of current designs are: wind instability, high cost of installations, and small power output of a single unit. The wind energy industry needs of revolutionary ideas to increase the capabilities of wind installations. This article suggests a revolutionary innovation which produces a dramatic increase in power per unit and is independent of prevailing weather and at a lower cost per unit of energy extracted. The main innovation consists of large free-flying air rotors positioned at high altitude for power and air stream stability, and an energy cable transmission system between the air rotor and a ground based electric generator. The air rotor system flies at high altitude up to 14 km. A stability and control is provided and systems enable the changing of altitude.

This article includes six examples having a high unit power output (up to 100 MW). The proposed examples provide the following main advantages: 1. Large power production capacity per unit – up to 5,000-10,000 times more than conventional ground-based rotor designs; 2. The rotor operates at high altitude of 1-14 km, where the wind flow is strong and steady; 3. Installation cost per unit energy is low. 4. The installation is environmentally friendly (no propeller noise).

---------------------

## Nomenclature (in metric system)

$A$ - front area of rotor [m$^2$];
$\alpha$ = 0.1 - 0.25 exponent of wind coefficient. One depends from Earth's surface roughness;
$A_a$ - wing area is served by aileron for balance of rotor (propeller) torque moment [m$^2$];
$A_w$ - area of the support wing [m$^2$];
$C$ - retail price of 1 kWh [$];
$c$ - production cost of 1 kWh [$];
$C_L$ - lift coefficient (maximum $C_L \approx 2.5$);
$C_D$ – drag coefficient;
$\Delta C_{L,a}$ - difference of lift coefficient between left and right ailerons;
$D$ – drag force [N];
$D_r$ - drag of rotor [N];
$E$ - annual energy produced by flow installation [J];
$F$ – annual profit [$];
$H_o$ = 10 m - standard altitude of ground wind installation [m];
$H$ - altitude [m];
$I$ - cost of Installation [$];
$K_1$ - life time (years);
$K_2$ – rotor lift coefficient (5-12 [kg/kW]);

*L* - length of cable [m];
$L_y$ – lift force of wing [N];
*M* – annual maintenance [$];
*N* – power [W, joule/sec];
$N_o$ - power at $H_o$ ;
*r* - distance from center of wing to center of aileron [m];
*R* - radius of rotor (turbine)[m];
*S* - cross-section area of energy transmission cable [m$^2$];
*V* - annual average wind speed [m/s];
$V_o$ - wind speed at standard altitude 10 m [m/s]($V_o$= 6 m/s);
*W* - weight of installation (rotor + cables)[kg];
$W_y$ – weight of cable [kg];
$\gamma$ - specific density of cable [kg/m$^3$];
$\eta$ - efficiency coefficient;
$\theta$ - angle between main (transmission) cable and horizontal surface;
$\lambda$ - ratio of blade tip speed to wind speed;
*v* - speed of transmission cable [m/s];
$\rho$ - density of flow, $\rho$ =1.225 kg/m$^3$ for air at sea level altitude *H* = 0; $\rho$ =0.736 at altitude *H* =5 km;
  $\rho$ = 0.413 at *H* =10 km;
$\sigma$ - tensile stress of cable [N/m$^2$].

## Introduction

Wind is a clean and inexhaustible source of energy that has been used for many centuries to grind grain, pump water, propel sailing ships, and perform other work.

Wind farm is the term used for a large number of wind machines clustered at a site with persistent favorable winds, generally near mountain passes. Wind farms have been erected in New Hampshire, in the Tehachapi Mountains. at Altamont Pass in California, at various sites in Hawaii, and may other locations. Machine capacities range from 10 to 500 kilowatts. In 1984 the total energy output of all wind farms in the United States exceeded 150 million kilowatt-hours.

A program of the United States Department of Energy encouraged the development of new machines, the construction of wind farms, and an evaluation of the economic effect of large-scale use of wind power.

The utilization of renewable energy ('green' energy) is currently on the increase. For example, a lot of wind turbines are being installed along the British coast. In addition, the British government has plans to develop off-shore wind farms along their coast in an attempt to increase the use of renewable energy sources. A total of $2.4 billion was injected into renewable energy projects over the last three years in an attempt to meet the government's target of using renewable energy to generate 10% of the country's energy needs by 2010.

This British program saves the emission of almost a millions tons of carbon dioxide. Denmark plans to get about 30% of their energy from wind sources.

Unfortunately, current wind energy systems have deficiencies which limit their commercial applications:
1. Wind energy is unevenly distributed and has relatively low energy density. Huge turbines cannot be placed on the ground, many small turbines must be used instead. In California, there are thousands of small wind turbines. However, while small turbines are relatively inefficient, very huge turbines placed at ground are also inefficient due to the relatively low wind energy density and their high cost. The current cost of wind energy is higher then energy of thermal power stations.

2. Wind power is a function of the cube of wind velocity. At surface level, wind has low speed and it is non-steady. If wind velocity decreases in half, the wind power decreases by a factor of 8 times.
3. The productivity of a wind-power system depends heavily on the prevailing weather.
4. Wind turbines produce noise and visually detract from the landscape.

There are many research programs and proposals for the wind driven power generation systems, however, all of them are ground or tower based. System proposed in this article is located at high altitude (up to the stratosphere), where strong permanent and steady streams are located. The also proposes a solution to the main technologist challenge of this system; the transfer of energy to the ground via a mechanical transmission made from closed loop, modern composite fiber cable.

The reader can find the information about this idea in [1], the wind energy in references [2]-[3], a detailed description of the innovation in [4]-[5], and new material used in the proposed innovation in [6]-[9]. The application of this innovation and energy transfer concept to other fields can be found in [10]-[19].

## Description of Innovation

Main proposed high altitude wind system is presented in fig.1. That includes: rotor (turbine) 1, support wing 2, cable mechanical transmission and keep system 3, electro-generator 4, and stabilizer 5. The transmission system has three cables (fig.1e): main (central) cable, which keeps the rotor at a given altitude, and two transmission mobile cables, which transfer energy from the rotor to the ground electric generator. The device of fig.1f allows changing a cable length and a rotor altitude. In calm weather the rotor can be support at altitude by dirigible 9 (fig.1c) or that is turned in vertical position and support by rotation from the electric generator (fig.1d). If the wind is less of a minimum speed for support of rotor at altitude the rotor may be supported by autogiro mode in position of fig.1d. The probability of full wind calm at a high altitude is small and depends from an installation location.

Fig.2 shows other design of the proposed high altitude wind installation. This rotor has blades, 10, connected to closed-loop cables. The forward blades have a positive angle and lift force. When they are in a back position the lift force equals zero. The rotor is supported at the high altitude by the blades and the wing 2 and stabilizer 5. That design also has energy transmission 3 connected to the ground electric generator 4.

Fig.3. shows a parachute wind high altitude installation. Here the blades are changed by parachutes. The parachutes have a large air drag and rotate the cable rotor 1. The wind 2 supports the installation in high altitude. The cable transmission 3 passes the rotor rotation to the ground electric generator 4.

A system of fig.4 uses a large Darries air turbine located at high altitude. This turbine has four blades. The other components are same with previous projects.

Wind turbine of fig.5 is a wind ground installation. Its peculiarity is a gigantic cable-blade rotor. That has a large power for low ground wind speed. It has four columns with rollers and closed-loop cable rotor with blades 10. The wind moves the blades, the blades move the cable, and the cable rotates an electric generator 4.

## Problems of launch, start, guidance, control, stability, and others

*Launching.* It is not difficult to launch the installations having support wing or blades as described in fig.1-4. If the wind speed is more than the minimum required speed (>2-3 m/s), the support wing lifts the installation to the desired altitude.

*Starting.* All low-speed rotors are self-starting. All high-speed rotors (include the ground rotor of fig.5) require an initial starting rotation from the ground motor-generator 4 (figs.1,5)**.**

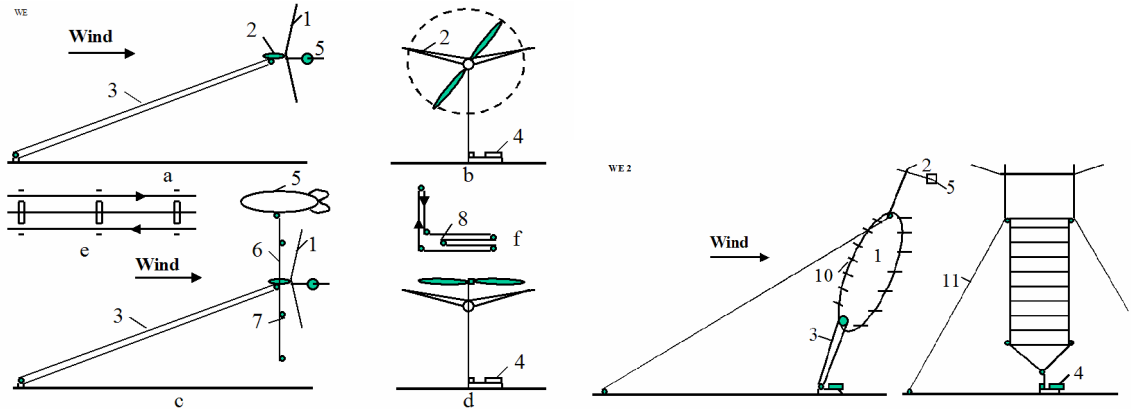

**Fig.1** (left). Propeller high altitude wind energy installation and cable energy transport system. Notation: a – side view; 1 – wind rotor; 2 – wing with ailerons; 3 – cable energy transport system; 4 – electric generator; 5 – stabilizer; b – front view; c – side view with a support dirigible 9, vertical cable 6, and wind speed sensors 7; d - keeping of the installation at a high altitude by rotate propeller; e – three lines of the transmission - keeper system. That includes: main (central) cable and two mobile transmission cables; f – energy transport system with variable altitude; 8 – mobile roller.

**Fig.2** (right). High altitude wind energy installation with the cable turbine. Notation: 10 – blades; 11 – tensile elements (bracing)(option).

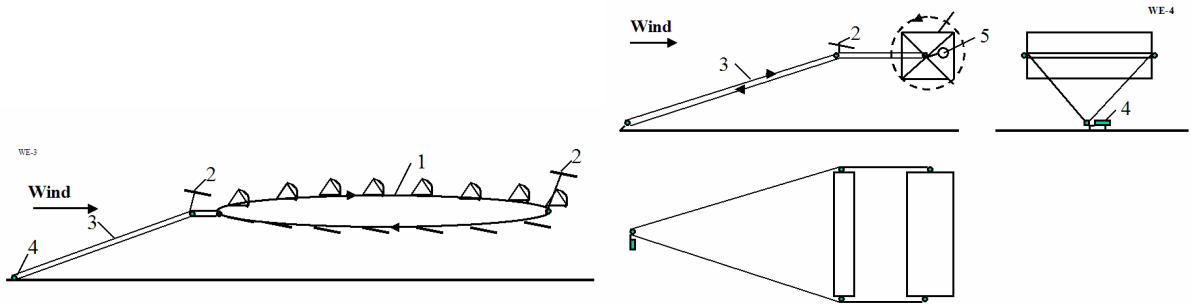

**Fig.3** (left). High altitude wind energy installation with the parachute turbine.
**Fig.4** (right). High altitude wind energy installation with Darrieus turbine.

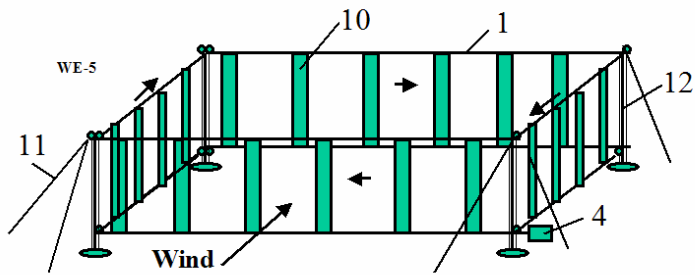

Fig.5. Ground wind cable rotor of a large power.

*Guidance and Control.* The control of power, revolutions per minute, and torque moment are operated by the turning of blades around the blade longitudinal axis. The control of altitude may be manual or automatic when the wind speed is normal and over admissible minimum. Control is effected by wing flaps and stabilizer (elevator), fin, and ailerons (figs. 1,2,4).

*Stability*. Stability of altitude is produced by the length of the cable. Stability around the blade longitudinal axis is made by stabilizer (see figs.1,2,4). Rotor directional stability in line with the flow can be provided by fins (figs. 1). When the installation has the support wing rigidly connected to the rotor, the stability is also attained by the correct location of the center of gravity of the installation (system rotor-wing) and the point of connection of the main cable and the tension elements. The center-of-gravity and connection point must be located within a relatively narrow range 0.2-0.4 of the average aerodynamic chord of the support wing (for example, see fig. 1). There is the same requirement for the additional support wings such as fig.2-4.

*Torque moment* is balanced by transmission and wing ailerons (see figs.1-4).

*The wing lift force, stress* of main cable are all regulated automatic by the wing flap or blade stabilizer.

***The location of the installation of fig.2 at a given point in the atmosphere*** may be provided by tension elements shown on fig.2. These tension elements provide a turning capability for the installation of approximately $\pm 45^0$ degrees in the direction of flow (see. Fig.2.).

*Minimum wind speed*. The required minimum wind-speed for most of the suggested installation designs is about 2 m/s. The probability of this low wing speed at high altitude is very small (less 0.001). This minimum may be decreased still further by using the turning propeller in an autogiro mode. If the wind speed is approximately zero, the rotor can be supported in the atmosphere by a balloon (dirigible) as is shown on fig.1c or a propeller rotated by the ground power station as is shown on fig.1d. The rotor system may also land on the ground and start again when the wind speed attains the minimum speed for flight.

*A Gusty winds*. Large pulsations of wind (aerodynamic energy) can be smoothed out by inertial fly-wheels.

The suggested Method and Installations for utilization of wind energy has following peculiarities from current conventional methods and installations:

1. Proposed installation allows the collection of energy from a large area – tens and even hundreds of times more than conventional wind turbines. This is possible because an expensive tower is not needed to fix our rotor in space. Our installation allows the use of a rotor with a very large diameter, for example 100-200 meters or more.
2. The proposed wind installations can be located at high altitude 100 m - 14 km. The wind speeds are 2-4 times faster and more stable at high altitude compared to ground surface winds used by the altitude of conventional windmills (10-70 meters of height). In certain geographic areas high altitude wind flows have a continuous or permanent nature. Since wind power increases at the cube of wind speed, wind rotor power increases by 27 times when wind speed increases by 3 times.
3. In proposed wind installation the electric generator is located at ground. There are proposals where electric generator located near a wind rotor and sends electric current to a ground by electric wares. However, our rotor and power are very large (see projects below). Proposed installations produce more power by thousands of times compared to the typical current wind ground installation (see point 1, 2 above). The electric generator of 20 MW weighs about 100 tons (specific weigh of the conventional electric generator is about 3-10 kg/kW). It is impossible to keep this weigh by wing at high altitude for wind speed lesser then 150 m/s.
4. One of the main innovations of the given invention is the **cable transfer** (transmission) of energy from the wind rotor located at high altitude to the electric generator located on ground. In proposed Installation it is used a new cable transmission made from artificial fibers. This transmission has less a weigh in thousands times then copper electric wires of equal power. The wire having diameter more 5 mm passes 1-2 ampere/sq.mm. If the electric generator produces 20 MW with voltage 1000 Volts, the wire cross-section area must be 20,000 mm$^2$, (wire diameter is160 mm). The cross-section area of the cable transmission of equal power is only 37 mm$^2$ (cable diameter 6.8 mm$^2$ for cable speed 300 m/s and admissible stress 200 kg/mm$^2$, see Project 1). The specific weight of copper is 8930 kg/m$^3$, the specific weight of artificial fibers is 1800 kg/m$^3$. If the cable length for altitude 10 km is 25 km the double copper wire weighs 8930 tons (!!), the fiber transmission cable weighs only 3.33 tons. It means the offered cable transferor energy of equal length is easier in 2682 times, then copper wire. The copper wires is very expensive, the artificial fiber is cheap.

All previous attempts to place the generator near the rotor and connect it to ground by electric transmission wires were not successful because the generator and wires are heavy.

## Some information about wind energy

The power of a wind engine strongly depends on the wind speed (to the third power). Low altitude wind ($H$ = 10 m) has the standard average speed $V$ = 6 m/s. High altitude wind is powerful and that has another important advantage, it is stable and constant. This is true practically everywhere.

Wind in the troposphere and stratosphere are powerful and permanent. For example, at an altitude of 5 km, the average wind speed is about 20 M/s, at an altitude 10-12 km the wind may reach 40 m/s (at latitude of about 20-35$^0$N).

There are permanent jet streams at high altitude. For example, at $H$ = 12-13 km and about 25$^0$N latitude. The average wind speed at its core is about 148 km/h (41 m/s). The most intensive portion, with a maximum speed 185 km/h (51 m/s) latitude 22$^0$, and 151 km/h (42 m/s) at latitude 35$^0$ in North America. On a given winter day, speeds in the jet core may exceed 370 km/h (103 m/s) for a distance of several hundred miles along the direction of the wind. Lateral wind shears in the direction normal to the jet stream may be 185 km/h per 556 km to right and 185 km/h per 185 km to the left.

The wind speed of $V$ = 40 m/s at an altitude $H$ = 13 km provides 64 times more energy than surface wind speeds of 6 m/s at an altitude of 10 m.

This is a gigantic renewable and free energy source. (See reference: *Science and Technolody, v.2, p.265*).

## Cable transmission energy problem

The primary innovations presented in this paper are locating the rotor at high altitude, and an energy transfer system using a cable to transfer mechanical energy from the rotor to a ground power station. The critical factor for this transfer system is the weight of the cable, and its air drag.

Twenty years ago, the mass and air drag of the required cable would not allow this proposal to be possible. However, artificial fibers are currently being manufactured, which have tensile strengths of 3-5 times more than steel and densities 4-5 times less then steel. There are also experimental fibers (whiskers) which have tensile strengths 30-100 times more than a steel and densities 2 to 5 times less than steel. For example, in the book [6] p.158 (1989), there is a fiber (whisker) **$C_D$**, which has a tensile strength of $\sigma$ = 8000 kg/mm$^2$ and density (specific gravity) of $\gamma$ = 3.5 g/cm$^3$. If we use an estimated strength of 3500 kg/mm$^2$ ($\sigma$ =7·10$^{10}$ N/m$^2$, $\gamma$ = 3500 kg/m$^3$), then the ratio is $\gamma/\sigma$ = 0.1×10$^{-6}$ or $\sigma/\gamma$ = 10×10$^6$. Although the described (1989) graphite fibers are strong ($\sigma/\gamma$ = 10×10$^6$), they are at least still ten times weaker than theory predicts. A steel fiber has a tensile strength of 5000 MPA (500 kg/sq.mm), the theoretical limit is 22,000 MPA (2200 kg/mm$^2$)(1987); the polyethylene fiber has a tensile strength 20,000 MPA with a theoretical limit of 35,000 MPA (1987). The very high tensile strength is due to its nanotubes structure.

Apart from unique electronic properties, the mechanical behavior of nanotubes also has provided interest because nanotubes are seen as the ultimate carbon fiber, which can be used as reinforcements in advanced composite technology. Early theoretical work and recent experiments on individual nanotubes (mostly MWNT's, Multi Wall Nano Tubes) have confirmed that nanotubes are one of the stiffest materials ever made. Whereas carbon-carbon covalent bonds are one of the strongest in nature, a structure based on a perfect arrangement of these bonds oriented along the axis of nanotubes would produce an exceedingly strong material. Traditional carbon fibers show high strength and stiffness, but fall far short of the theoretical, in-plane strength of graphite layers by an order of magnitude. Nanotubes come close to being the best fiber that can be made from graphite.

For example, whiskers of Carbon nanotube (CNT) material have a tensile strength of 200 Giga-Pascals and a Young's modulus over 1 Tera Pascals (1999). The theory predicts 1 Tera Pascals and a Young's modules of 1-5 Tera Pascals. The hollow structure of nanotubes makes them very light (the specific density varies from 0.8 g/cc for SWNT's (Single Wall Nano Tubes) up to 1.8 g/cc for MWNT's, compared to 2.26 g/cc for graphite or 7.8 g/cc for steel).

Specific strength (strength/density) is important in the design of the systems presented in this paper; nanotubes have values at least 2 orders of magnitude greater than steel. Traditional carbon fibers have a specific strength 40 times that of steel. Since nanotubes are made of graphitic carbon, they have good resistance to chemical attack and have high thermal stability. Oxidation studies have shown that the onset of oxidation shifts by about $100^0$ C or higher in nanotubes compared to high modulus graphite fibers. In a vacuum, or reducing atmosphere, nanotube structures will be stable to any practical service temperature.

The artificial fibers are cheap and widely used in tires and everywhere. The price of SiC whiskers produced by Carborundum Co. with $\sigma$=20,690 MPa and $\gamma$=3.22 g/cc was $440 /kg in 1989. The market price of nanotubes is too high presently (~$200 per gram)(2000). In the last 2-3 years, there have been several companies that were organized in the US to produce and market nanotubes. It is anticipated that in the next few years, nanotubes will be available to consumers for less than $100/pound.

Below, the author provides a brief overview of recent research information regarding the proposed experimental (tested) fibers. In addition, the author also addresses additional examples, which appear in these projects and which can appear as difficult as the proposed technology itself. The author is prepared to discuss the problems with organizations which are interested in research and development related projects.

**Table # 1**. Material properties.

| Material Whiskers | Tensile strength kg/mm$^2$ | Density g/cc | Fibers | Tensile strength kg/mm$^2$ | Density g/cc |
|---|---|---|---|---|---|
| AlB$_{12}$ | 2650 | 2.6 | QC-8805 | 620 | 1.95 |
| B | 2500 | 2.3 | TM9 | 600 | 1.79 |
| B$_4$C | 2800 | 2.5 | Thorael | 565 | 1.81 |
| TiB$_2$ | 3370 | 4.5 | Allien 1 | 580 | 1.56 |
| SiC | 1380-4140 | 3.22 | Allien 2 | 300 | 0.97 |

Reference [6]-[9].

Industrial fibers with $\sigma$ = 500-600 kg/mm$^2$, $\gamma$ = -1800 kg/m$^3$, and $\sigma/\gamma$ = 2,78x10$^6$ are used in all our projects (admissible $\sigma$ =200-250 kg/mm$^2$)(see below).

## Brief Theory of Estimation of Suggested Installations
### Rotor

Power of a wind energy $N$ [Watt, Joule/sec]

$$N=0.5\eta\rho AV^3 \quad [W] \ . \qquad (1)$$

The coefficient of efficiency, $\eta$, equals 0.15-0.35 for low speed rotors (ratio of blade tip speed to wind speed equals $\lambda \approx 1$); $\eta$ = 0.35-0.5 for high speed rotors ($\lambda$ = 5-7). The Darrieus rotor has $\eta$ = 0.35 - 0.4. The propeller rotor has $\eta$ = 0.45-0.50. The theoretical maximum equals $\eta$ = 0.67.

The energy is produced in one year is (1 year $\approx 30.2\times10^6$ work sec) [J]

$$E=3600\times24\times350 \approx 30\times10^6 N \quad [J]. \qquad (1')$$

Wind speed increases with altitude as follows
$$V=(H/H_o)^\alpha V_o , \qquad (2)$$
where $\alpha = 0.1 - 0.25$ exponent coefficient depends from surface roughness. When the surface is water, $\alpha = 0.1$; when surface is shrubs and woodlands $\alpha = 0.25$.

Power increases with altitude as the cube of wind speed
$$N=(H/H_o)^{3\alpha} N_o , \qquad (3)$$
where $N_o$ is power at $H_o$.

The drag of the rotor equals
$$D_r = N/V . \qquad (4)$$

The lift force of the wing, $L_y$, is
$$L_y = 0.5 C_L \rho V^2 A_w , \quad L_y \approx W, \qquad (5)$$
where $C_L$ is lift coefficient (maximum $C_L \approx 2.5$), $A_w$ is area of the wing, W is weight of installation + 0.5 weight of all cables.

The drag of the wing is
$$D = 0.5 C_D \rho V^2 A_w , \qquad (6)$$
where $C_D$ is the drag coefficient (maximum $C_D \approx 1.2$).

The optimal speed of the parachute rotor equals $1/3 V$ and the theoretical maximum of efficiency coefficient is 0.5.

The annual energy produced by the wind energy extraction installation equals
$$E = 8.33 N \quad [kWh]. \qquad (7)$$

**Cable Energy Transfer, Wing Area, and other Parameters**

Cross-section area of transmission cable, $S$, is
$$S = N/v\sigma, \qquad (8)$$
Cross-section area of main cable, $S_m$, is
$$S_m = (D_r + D)/\sigma, \qquad (8')$$
Weight of cable is
$$W_r = SL\gamma , \qquad (9)$$
The production cost, $c$, in kWh is
$$c = \frac{M + I/K_1}{E} , \qquad (10)$$
The annual profit
$$F = (C-c)E . \qquad (11)$$
The required area of the support wing is
$$A_w = \frac{\eta A \sin\theta}{C_L} , \qquad (12)$$
where $\theta$ is the angle between the support cable and horizontal surface.

The wing area is served by ailerons for balancing of the rotor (propeller) torque moment
$$A_a = \frac{\eta A R}{\lambda_i \Delta C_{L,a} r} , \qquad (13)$$

The minimum wind speed for installation support by the wing alone
$$V_{min} = \sqrt{\frac{2W}{C_{L,max} \rho A_w}} , \qquad (14)$$
where $W$ is the total weight of the airborne system including transmission. If a propeller rotor is used in a gyroplane mode, minimal speed will decrease by 2-2,5 times. If wind speed equals zero, the required power for driving the propeller in a propulsion (helicopter) mode is

$$N_s = W/K_2 \quad [kW], \quad (15)$$

The specific weight of energy storage (flywheel) can be estimated by

$$E_s = \sigma/2\gamma \quad [J/kg]. \quad (16)$$

For example, if $\sigma = 200$ kg/mm$^2$, $\gamma = 1800$ kg/m$^3$, then $\mathbf{E_s} = 0.56$ MJ/*kg or* $\mathbf{E_s} = 0.15$ kWh/kg.

For comparison of the different ground wind installations their efficiency and parameters are computed for the standard wind conditions: the wind speed equals $V=6$ m/s at the altitude $H=10$ m.

# *Projects*
## Project 1
### High-speed air propeller rotor (fig.1)

For example, let us consider a rotor diameter of 100 m ($A = 7850$ m$^2$), at an altitude $H = 10$ km ($\rho = 0.4135$ kg/m$^3$), wind speed of $V = 30$ m/s, an efficiency coefficient of $\eta = 0.5$, and a cable tensile stress of $\sigma = 200$ kg/mm$^2$.

Then the power produced is $N = 22$ MW [Eq. (1)], which is sufficient for city with a population of 250,000. The rotor drag is $D_r = 73$ tons [Eq.(4)], the cross-section of the main cable area is $S = 1.4 D_r/\sigma$ =1.35×73/0.2 ≈ 500 mm$^2$, the cable diameter equals $d = 25$ mm; and the cable weight is $W = 22.5$ tons [Eq.(9)] (for $L = 25$ km). The cross-section of the transmission cable is $S = 36.5$ mm$^2$ [Eq.(8)], $d = 6.8$ mm, weight of two transmission cables is $W = 3.33$ tons for cable speed $v = 300$ m/s [Eq.(9)].

The required wing size is 20×100 m ($C_L = 0.8$) [Eq.(12)], wing area served by ailerons is 820 sq.m [Eq.(13)]. If $C_L = 2$, the minimum speed is 2 m/s [Eq.(14)].

The installation will produce an annual energy $E = 190$ GWh [Eq.(7)]. If the installation cost is $200K, has a useful life of 10 years, and requires maintenance of $50K per year, the production cost is $c = 0.37$ cent per kWh [Eq(10)]. If retail price is $0.15 per kWh, profit $0.1 per kWh, the total annual profit is $19 millions per year [Eq.(11)].

## The project #2
### Large air propeller at altitude H = 1 km (fig.1)

Let us consider a propeller diameter of 300 m, with an area $A = 7 \times 10^4$ m$^2$, at an altitude $H = 1$ km, and a wind speed of 13 m/s. The average blade tip speed is 78 m/s.

The full potential power of the wind streamer flow is 94.2 MW. If the coefficient of efficiency is 0.5 the useful power is $N = 47.1$ MW. For other wind speed. the useful power is: $V = 5$ m/s, $N = 23.3$ MW; $V = 6$ m/s, $N = 47.1$ MW; $V = 7$ m/s, $N = 74.9$ MW; $V = 8$ m/s, $N = 111.6$ MW; $V = 9$ m/s, $N = 159$ MW; $V = 10$ m/s, $N = 218$ MW.

### Estimation of economical efficiency

Let us assume that the cost of the Installation is $3 million, a useful life of 10 years, and request maintenance of $100,000/year. The energy produced in one year is $E = 407$ GWh [Eq.(7)]. The basic cost of energy is $0.01 /kWh.

### The some technical parameters
### Altitude H = 1 km

The drag is about 360 tons. Ground connection (main) cable has cross-section area of 1800 sq,mm [Eq.(8')], $d = 48$ mm, and has a weight of 6480 kg. The need wing area is 60x300 m. The aileron area requested for turbine balance is 6740 sq.m.

If the transmission cable speed is 300 m/s, the cross-section area of transmission cable is 76 sq.mm and the cable weight is 684 kg (composite fiber).

### Altitude H = 13 km

At an altitude of $H = 13$ km. the air density is $\rho = 0.2666$, and the wind speed is $V = 40$ m/s. The power for efficiency coefficient 0.5 is 301.4 MgW. The drag of the propeller is approximately 754 tons. The connection cable has a cross-sectional area of 3770 sq.mm, a diameter is $d = 70$ mm and a weight of 176 tons. The transmission cable has a sectional area 5 sq.c and a weight of 60 tons (vertical transmission only 12 tons).

The installation will produce energy $E$=2604 GWh per year. If the installation costs $5 million, maintenance is $200,000/year, and the cost of 1 kWh will be $0.0097/kWh.

# Project #3

### Air low speed wind engine with free flying cable flexible rotor (fig.2)

Let us consider the size of cable rotor of width 50 m, a rotor diameter of 1000 m, then the rotor area is $A$=50×1000=50,000 sq.m. The angle rope to a horizon is 70°. The angle of ratio lift/drag is about 2.5°.

The average conventional wind speed at an altitude $H$ = 10 m is $V$ = 6 m/s. It means that the speed at the altitude 1000 m is 11.4 - 15 m/s. Let us take average wind speed $V$=13 m/s at an altitude $H$ = 1 km.
The power of flow is

$$N=0.5\rho V^3 A \cos 20^0 = 0.5 \times 1.225 \times 13^3 \times 1000 \times 50 \times 0.94 = 63 \text{ MW}.$$

If the coefficient efficiency is $\eta$ = 0.2 the power of installation is

$$\eta = 0.2 \times 63 = 12.5 \text{ MW}.$$

The energy 12.5 MW is enough for a city with a population at 150,000.
If we decrease our Installation to a 100x2000 m the power decreases approximately by 6 times (because the area decreases by 4 times, wind speed reaches more 15 m/s at this altitude. Power will be 75 MW. This is enough for a city with a population about 1 million of people.

If the average wind speed is different for given location the power for the basis installation will be: $V$ =5 m/s, $N$ =7.25 MW; $V$ =6 m/s, $N$ =12.5 MW; $V$=7 m/s, $N$ = 19.9 MW; $V$ = 8 m/s, $N$ = 29,6 MW; $V$ = 9 m/s, $N$ = 42.2 MW; $V$ = 10 m/s, $N$ = 57.9 MW.

### Economical efficiency

Let us assume that the cost of our installation is $1 million. According to the book "Wind Power" by P. Gipe [2], the conventional wind installation with the rotor diameter 7 m costs $20,000 and for average wind speeds of 6 m/s has power 2.28 kW, producing 20,000 kWh per year. To produce the same amount of power as our installation using by conventional methods, we would need 5482 (12500/2.28) conventional rotors, costing $110 million.  Let us assume that our installation has a useful life of 10 years and a maintenance cost is $50,000/year. Our installation produces 109,500,000 kWh energy per year. Production costs of energy will be approximately 150,000/109,500,000 = 0.14 cent/kWh. The retail price of 1 kWh of energy in New York City is $0.15 now. The revenue is 16 millions. If profit from 1 kWh is $0.1, the total profit is more 10 millions per year.

**Estimation some technical parameters**

The cross-section of main cable for an admissible fiber tensile strange $\sigma$ = 200kg/sq.mm is $S$ =2000/0.2 = 10,000 mm$^2$. That is two cable of diameter $d$ =80 mm. The weight of the cable for density 1800 kg/m$^3$ is

$$W = SL\gamma = 0.01 \times 2000 \times 1800 = 36 \text{ tons}.$$

Let us assume that the weight of 1 sq.m of blade is 0.2 kg/m$^2$ and the weight of 1 m of bulk is 2 kg. The weight of the 1 blade will be 0.2 x 500 = 100 kg, and 200 blades are 20 tons. If the weight of one bulk is 0.1 ton, the weight of 200 bulks is 20 tons.

The total weight of main parts of the installation will be 94 tons. We assume 100 tons for purposes of our calculations.

The minimum wind speed when the flying rotor can supported in the air is (for $C_y$ = 2)

$$V=(2W/C_y\rho S)^{0.5}=(2\times 100 \times 10^4/2 \times 1.225 \times 200 \times 500)^{0.5} = 2.86 \text{ m/s}$$

The probability of the wind speed falling below 3 m/s when the average speed is 12 m/s, is zero, and for 10 m/s is 0.0003. This equals 2.5 hours in one year, or less than one time per year. The wind at high altitude has greater speed and stability than near ground surface. There is a strong wind at high altitude even when wind near the ground is absent. This can be seen when the clouds move in a sky on a calm day.

# Project #4
## Low speed air drag rotor (fig.3)

Let us consider a parachute with a diameter of 100 m, length of rope 1500 m, distance between the parachutes 300 m, number of parachute 3000/300 = 10, number of worked parachute 5, the area of one parachute is 7850 sq.m, the total work area is $A$ = 5 x 7850 = 3925 sq.m. The full power of the flow is 5.3 MW for $V$=6 m/s. If coefficient of efficiency is 0.2 the useful power is $N$ = 1 MW. For other wind speed the useful power is: $V$ =5 m/s, $N$ =0.58 MW; $V$ =6 m/s, $N$ = 1 MW; $V$ =7 m/s, $N$ =1.59 MW; $V$ = 8 m/s, $N$=2.37 MW; $V$ = 9 m/s, $N$ =3.375 MW; $V$ = 10 m/s, $N$ = 4.63 MW.

### Estimation of economical efficiency

Let us take the cost of the installation $0.5 million, a useful life of 10 years and maintenance of $20,000/year. The energy produced in one year (when the wind has standard speed 6 m/s) is $E$ = 1000x24x360 = 8.64 million kWh. The basic cost of energy is 70,000/8,640,000 = 0.81 cent/kWh.

### The some technical parameters

If the thrust is 23 tons, the tensile stress is 200 kg/sq.mm (composed fiber), then the parachute cable diameter is 12 mm, The full weight of the installation is 4.5 tons. The support wing has size 25x4 m.

# Project #5
## High speed air Darreus rotor at an altitude 1 km (fig.4)

Let us consider a rotor having the diameter of 100 m, a length of 200 m (work area is 20,000 sq.m). When the wind speed at an altitude $H$=10 m is $V$ =6 m/s, then at an altitude $H$ = 1000 m it is 13 m/s. The full wind power is 13,46 MW. Let us take the efficiency coefficient 0.35, then the power of the Installation will be $N$ = 4.7 MW. The change of power from wind speed is: $V$ = 5 m/s, $N$ = 2.73 MW; $V$ = 6 m/s, $N$ = 4.7 MW; $V$ = 7 m/s, $N$ = 7.5 MW; $V$ = 8 m/s, $N$ = 11.4 MW; $V$ = 9m/s, $N$ = 15.9 MW; $V$ = 10 m/s, $N$ = 21.8 MW.

At an altitude of $H$ = 13 km with an air density 0.267 and wind speed $V$ = 40 m/s, the given installation will produce power $N$ = 300 MW.

### Estimation of economical efficiency

Let us take the cost of the Installation at $1 million, a useful life of 10 years, and maintenance of $50,000 /year. Our installation will produce $E$ = 41 millions kWh per year (when the wind speed equals 6 m/s at an altitude 10 m). The prime cost will be 150,000/41,000,000 = 0.37 cent/kWh. If the customer price is $0.15/kWh and profit from 1 kWh is $0.10 /kWh the profit will be $4.1 million per year.

### Estimation of technical parameters

The blade speed is 78 m/s. Numbers of blade is 4. Number of revolution is 0.25 revolutions per second. The size of blade is 200x0.67 m. The weight of 1 blade is 1.34 tons. The total weight of the Installation is about 8 tons. The internal wing has size 200x2.3 m. The additional wing has size 200x14.5 m and weight 870 kg. The cross-section area of the cable transmission having an altitude of $H$ = 1 km is 300 sq.mm, the weight is 1350 kg.

# Project #6
## Ground Wind High Speed Engine (fig.5)

Let us consider the ground wind installation (fig.5) with size 500x500x50 meters. The work area is 500x50x2= 50,000 sq.m. The tower is 60 meter tall, the flexible rotor located from 10 m to 60 m. If the wind speed at altitude 10 m is 6 m/s, that equals 7.3 m/s at altitude 40 m.

The theoretical power is

$$N_t = 0.5\rho V^3 A = 0.5 \times 1.225 \times 7.3^3 \times 5 \times 10^4 = 11.9 \text{ MgW}.$$

For coefficient of the efficiency equals 0.45 the useful power is

$$N = 0.45 \times 11.9 = 5.36 \text{ MW}.$$

For other wind speed at an altitude 6 m/s the useful power is: $V$ = 5 m/s,  $N$ = 3.1 MW; $V$ = 6 m/s,  $N$ = 5.36 MW; $V$ = 7 m/s,  $N$ = 8.52 MW; $V$ = 8 m/s,  $N$ = 12.7 MW; $V$ = 9 m/s,  $N$ = 18.1 MW; $V$ = 10 m/s,  $N$ = 24.8 MW.

### Economic estimation

In this installation the rotor will be less expensive than previous installations because the high-speed rotor has a smaller number of blades and smaller blades (see technical data below). However this installation needs 4 high (60 m) columns. Take the cost of the installation at $1 million with a useful life of 10 years. The maintenance is projected at about $50,000 /year.

This installation will produce $E$ = 5360 kW x 8760 hours = 46.95 MWh energy (for the annual average wind-speed   $V$ = 6 m/s at $H$ = 10 m). The cost of 1 kWh is 150,000/46,950,000 = 0.4 cent/kWh. If the retail price is $0.15/kWh and delivery cost 30%, the profit is $0.10 per kWh, or $4.7 million per year.

### Estimation of some technical parameters

The blade speed is 6 x 7.3 = 44 m/s. The distance between blades is 44 m. The number of blade is 4000/44 = 92.

## Discussion and Conclusion

Conventional windmills are approached their maximum energy extraction potential relative to their installation cost. No relatively progress has been made in windmill technology in the last 50 years. The wind energy is free, but its production more expensive then its production in heat electric stations. Current wind installations cannot essential decrease a cost of kWh, stability of energy production. They cannot increase of power of single energy unit. The renewable energy industry needs revolutionary ideas that improve performance parameters (installation cost and power per unit) and that significantly decreases (in 10-20 times) the cost of energy production. This paper offers ideas that can move the wind energy industry from stagnation to revolutionary potential.

The following is a list of benefits provided by the proposed system compared to current installations:

1. The produced energy at least in 10 times cheaper then energy received of all conventional electric stations includes current wind installation.
2. The proposed system is relatively inexpensive (no expensive tower), it can be made with a very large thus capturing wind energy from an enormous area (hundreds of times more than typical wind turbines).
3. The power per unit of proposed system in some hundreds times more of typical current wind installations.
4. The proposed installation not requires large ground space.
5. The installation may be located near customers and not require expensive high voltage equipment. It is not necessary to have long, expensive, high-voltage transmission lines and substations. Ocean going vessels can use this installation for its primary propulsion source.
6. No noise and bad views.
7. The energy production is more stability because the wind is steadier at high altitude. The wind may be zero near the surface but it is typically strong and steady at higher altitudes. This can be observed when it is calm  on the ground, but clouds are moving in the sky. There are a strong permanent air streams at a high altitude at many regions of the USA.
8. The installation can be easy relocated in other place.

As with any new idea, the suggested concept is in need of research and development. The theoretical problems do not require fundamental breakthroughs. It is necessary to design small, free flying installations to study and get an experience in the design, launch, stability, and the cable energy transmission from a flying wind turbine to a ground electric generator.

This paper has suggested some design solutions from patent application [4]. The author has many detailed analysis in addition to these presented projects. Organizations interested in these projects can address the author (http://Bolonkin.narod.ru , aBolonkin@juno.com , abolonkin@gmail.com).

The other ideas are in [11]-[40].

## References

(The reader finds some of this articles in http://Bolonkin.narod.ru/p65.htm and http://arxiv.org)

1. Bolonkin A.A., Utilization of Wind Energy at High Altitude, AIAA-2004-5756, AIAA-2004-5705. International Energy Conversion Engineering Conference at Providence, RI, USA, Aug.16-19, 2004.
2. Gipe P., Wind Power, Chelsea Green Publishing Co., Vermont, 1998.
3. Thresher R.W. and etc, Wind Technology Development: Large and Small Turbines, NRFL, 1999.
4. Bolonkin, A.A., "Method of Utilization a Flow Energy and Power Installation for It", USA patent application 09/946,497 of 09/06/2001.
5. Bolonkin, A.A., Transmission Mechanical Energy to Long Distance. AIAA-2004-5660.
6. Galasso F.S., Advanced Fibers and Composite, Gordon and Branch Scientific Publisher, 1989.
7. Carbon and High Performance Fibers Directory and Data Book, London-New. York: Chapmen& Hall, 1995, 6$^{th}$ ed., 385 p.
8. Concise Encyclopedia of Polymer Science and Engineering, Ed. J.I.Kroschwitz, N. Y.,Wiley, 1990, 1341 p.
9. Dresselhaus, M.S., Carbon Nanotubes, by, Springer, 2000.
10. Bolonkin, A.A., "Inexpensive Cable Space Launcher of High Capability", IAC-02-V.P.07, 53$^{rd}$ International Astronautical Congress. The World Space Congress – 2002, 10-19 Oct. 2002/Houston, Texas, USA. JBIS, Vol.56, pp.394-404, 2003.
11. Bolonkin, A.A, "Non-Rocket Missile Rope Launcher", IAC-02-IAA.S.P.14, 53$^{rd}$ International Astronautical Congress. The World Space Congress – 2002, 10-19 Oct 2002/Houston, Texas, USA. JBIS, Vol.56, pp.394-404, 2003.
12. Bolonkin, A.A., "Hypersonic Launch System of Capability up 500 tons per day and Delivery Cost $1 per Lb". IAC-02-S.P.15, 53$^{rd}$ International Astronautical Congress. The World Space Congress – 2002, 10-19 Oct 2002/Houston, Texas, USA. JBIS, Vol.57, pp.162-172. 2004.
13. Bolonkin, A.A., "Employment Asteroids for Movement of Space Ship and Probes". IAC-02-S.6.04, 53$^{rd}$ International Astronautical Congress. The World Space Congress – 2002, 10-19 Oct. 2002/Houston, USA. JBIS, Vol.56, pp.98-197, 2003.
14. Bolonkin, A.A., "Non-Rocket Space Rope Launcher for People", IAC-02-V.P.06, 53$^{rd}$ International Astronautical Congress. The World Space Congress – 2002, 10-19 Oct 2002/Houston, Texas, USA. JBIS, Vol.56, pp.231-249, 2003.
15. Bolonkin, A.A., "Optimal Inflatable Space Towers of High Height". COSPAR-02 C1.1-0035-02, 34$^{th}$ Scientific Assembly of the Committee on Space Research (COSPAR). The World Space Congress – 2002, 10-19 Oct 2002/Houston, Texas, USA. JBIS, Vol.56, pp.87-97, 2003.
16. Bolonkin, A.A., "Non-Rocket Earth-Moon Transport System", COSPAR-02 B0.3-F3.3-0032-02, 34$^{th}$ Scientific Assembly of the Committee on Space Research (COSPAR). The World Space Congress – 2002, 10-19 Oct 2002, Houston, Texas, USA. "Advanced Space Research", Vol.31, No. 11, pp. 2485-2490, 2003.
17. Bolonkin, A.A., "Non-Rocket Earth-Mars Transport System", COSPAR-02B0.4-C3.4-0036-02, 34$^{th}$ Scientific Assembly of the Committee on Space Research (COSPAR). The World Space Congress – 2002, 10-19 Oct 2002/Houston, Texas, USA. *Actual problems of aviation and space system*. No.1(15), vol.8, pp.63-73, 2003.
18. Bolonkin, A.A., "Transport System for delivery Tourists at Altitude 140 km". IAC-02-IAA.1.3.03, 53$^{rd}$ International Astronautical Congress. The World Space Congress – 2002, 10-19 Oct. 2002/Houston, Texas, USA. JBIS, Vol.56, pp.314-327, 2003.
19. Bolonkin, A.A., "Hypersonic Gas-Rocket Launch System.", AIAA-2002-3927, 38$^{th}$ AIAA/ASME/SAE/ ASEE Joint Propulsion Conference and Exhibit, 7-10 July, 2002. Indianapolis, IN, USA.
20. Bolonkin, A.A., Multi-Reflex Propulsion Systems for Space and Air Vehicles and Energy Transfer for Long Distance, JBIS, Vol, 57, pp.379-390, 2004.
21. Bolonkin A.A., Electrostatic Solar Wind Propulsion System, AIAA-2005-3653. 41 Propulsion Conference, 10-12 July, 2005, Tucson, Arizona, USA.
22. Bolonkin A.A., Electrostatic Utilization of Asteroids for Space Flight, AIAA-2005-4032. 41 Propulsion Conference, 10-12 July, 2005, Tucson, Arizona, USA.

23. Bolonkin A.A., Kinetic Anti-Gravitator, AIAA-2005-4504. 41 Propulsion Conference, 10-12 July, 2005, Tucson, Arizona, USA.
24. Bolonkin A.A., Sling Rotary Space Launcher, AIAA-2005-4035. 41 Propulsion Conference, 10-12 July, 2005, Tucson, Arizona, USA.
25. Bolonkin A.A., Radioisotope Space Sail and Electric Generator, AIAA-2005-4225. 41 Propulsion Conference, 10-12 July, 2005, Tucson, Arizona, USA.
26. Bolonkin A.A., Guided Solar Sail and Electric Generator, AIAA-2005-3857. 41 Propulsion Conference, 10-12 July, 2005, Tucson, Arizona, USA.
27. Bolonkin A.A., Problems of Electrostatic Levitation and Artificial Gravity, AIAA-2005-4465. 41 Propulsion Conference, 10-12 July, 2005, Tucson, Arizona, USA.
28. A.A. Bolonkin, Space Propulsion using Solar Wing and Installation for It. Russian patent application #3635955/23 126453, 19 August, 1983 (in Russian). Russian PTO.
29. A.A, Bolonkin, Installation for Open Electrostatic Field. Russian patent application #3467270/21 116676, 9 July, 1982 (in Russian). Russian PTO.
30. A.A.Bolonkin, Getting of Electric Energy from Space and Installation for It. Russian patent application #3638699/25 126303, 19 August, 1983 (in Russian). Russian PTO.
31. A.A.Bolonkin, Protection from Charged Particles in Space and Installation for It. Russian patent application #3644168 136270 of 23 September 1983, (in Russian). Russian PTO.
32. A.A.Bolonkin, Method of Transformation of Plasma Energy in Electric Current and Installation for It. Russian patent application #3647344 136681 of 27 July 1983 (in Russian), Russian PTO.
33. A.A.Bolonkin, Method of Propulsion using Radioisotope Energy and Installation for It. of Plasma Energy in Electric Current and Installation for it. Russian patent application #3601164/25 086973 of 6 June, 1983 (in Russian), Russian PTO.
34. A.A.Bolonkin, Transformation of Energy of Rarefaction Plasma in Electric Current and Installation for it. Russian patent application #3663911/25 159775 of 23 November 1983 (in Russian). Russian PTO.
35. A.A.Bolonkin, Method of a Keeping of a Neutral Plasma and Installation for it. Russian patent application #3600272/25 086993 of 6 june 1983 (in Russian). Russian PTO.
36. A.A.Bolonkin, Radioisotope Propulsion. Russian patent application #3467762/25 116952 of 9 July 1982 (in Russian). Russian PTO.
37. A.A.Bolonkin, Radioisotope Electric Generator. Russian patent application #3469511/25 116927 of 9 July 1982 (in Russian). Russian PTO.
38. A.A.Bolonkin, Radioisotope Electric Generator. Russian patent application #3620051/25 108943 of 13 July 1983 (in Russian). Russian PTO.
39. A.A.Bolonkin, Method of Energy Transformation of Radioisotope Matter in Electricity and Installation for it. Russian patent application #3647343/25 136692 of 27 July 1983 (in Russian). Russian PTO.
40. A.A.Bolonkin, Method of stretching of thin film. Russian patent application #3646689/10 138085 of 28 September 1983 (in Russian). Russian PTO.
41. Bolonkin, A.A. and R.B. Cathcart, Inflatable 'Evergreen' Dome Settlements for Earth's Polar Regions. Clean. Techn. Environ. Policy. DOI 10.1007/s10098.006-0073.4 .
42. Bolonkin, A.A. and R.B. Cathcart, "A Cable Space Transportation System at the Earth's Poles to Support Exploitation of the Moon", *Journal of the British Interplanetary Society* 59: 375-380, 2006.
43. Bolonkin A.A., Cheap Textile Dam Protection of Seaport Cities against Hurricane Storm Surge Waves, Tsunamis, and Other Weather-Related Floods, 2006. http://arxiv.org.
44. Bolonkin, A.A. and R.B. Cathcart, Antarctica: A Southern Hemisphere Windpower Station? Arxiv, 2007.
45. Bolonkin A.A., Cathcart R.B., Inflatable 'Evergreen' Polar Zone Dome (EPZD) Settlements, 2006. http://arxiv.org
46. Bolonkin, A.A. and R.B. Cathcart, The Java-Sumatra Aerial Mega-Tramway, 2006. http://arxiv.org.
47. Bolonkin, A.A., "Optimal Inflatable Space Towers with 3-100 km Height", *Journal of the British Interplanetary Society* Vol. 56, pp. 87 - 97, 2003.
48. Bolonkin A.A., *Non-Rocket Space Launch and Flight, Elsevier*, London, 2006, 488 ps.
49. Macro-Engineering: *A Challenge for the Future*. Springer, 2006. 318 ps. Collection articles.
50. Cathcart R.B. and Bolonkin, A.A. Ocean Terracing, 2006. http://arxiv.org.